\documentclass[11pt]{article}
\usepackage{PRIMEarxiv}
\usepackage[round]{natbib}

\usepackage[utf8]{inputenc} % allow utf-8 input
\usepackage[T1]{fontenc}    % use 8-bit T1 fonts
\usepackage[linkcolor=blue,allcolors=blue]{hyperref}
\usepackage{url}            % simple URL typesetting
\usepackage{booktabs}       % professional-quality tables
\usepackage{amsfonts}       % blackboard math symbols
\usepackage{nicefrac}       % compact symbols for 1/2, etc.
\usepackage{microtype,nicefrac}      % microtypography
\usepackage{xspace}
\usepackage{natbib}
\usepackage{pifont}

\usepackage{comment}
\usepackage{amsmath}
\usepackage{graphicx}
\usepackage{rotating}

\usepackage{amssymb}
\usepackage{autobreak}
\usepackage{mathtools}
\usepackage[dvipsnames]{xcolor}
\usepackage{bbm}
\usepackage{algorithm}
\usepackage[parfill]{parskip}
\usepackage{algpseudocode}
\usepackage[export]{adjustbox}% http://ctan.org/pkg/adjustbox
\usepackage{subcaption}
\usepackage[toc]{appendix}
\usepackage{minitoc}
\makeatletter
\usepackage[flushleft]{threeparttable}

\title{Insights on the dip of fault zones in Southern California from modeling of seismicity with anisotropic point processes}

\author{Zachary E. Ross \\
Seismological Laboratory, California Institute of Technology, Pasadena, CA, USA 91125}
\begin{document}

	\maketitle
		\begin{abstract}
        Accurate models of fault zone geometry are important for scientific and hazard applications. While seismicity can provide high-resolution point measurements of fault geometry, extrapolating these measurements to volumes may involve making strong assumptions. This is particularly problematic in distributed fault zones, which are commonly observed in immature faulting regions. In this study, we focus on characterizing the dip of fault zones in Southern California with the goal of improving fault models. We introduce a novel technique from spatial point process theory to quantify the orientation of persistent surficial features in seismicity, even when embedded in wide shear zones. The technique makes relatively mild assumptions about fault geometry and is formulated with the goal of determining the dip of a fault zone at depth. The method is applied to 11 prominent seismicity regions in Southern California. Overall, the results compare favorably with the geometry models provided by the SCEC Community Fault Model and other focused regional studies. More specifically, we find evidence that the Southern San Andreas and San Jacinto fault zones are both northeast dipping at seismogenic depths at the length scales of 1.0-4.0 km. In addition, we find more limited evidence for some depth dependent variations in dip that suggest a listric geometry. The developed technique can provide an independent source of information from seismicity to augment existing fault geometry models.
		\end{abstract}
    
        \section{Introduction}
        The geometrical properties of fault zones are basic, yet fundamental quantities in earthquake science. Earthquake rupture simulations need fault geometry models that faithfully capture these attributes in order to adequately quantify expected seismic hazard with physics-based approaches \citep{shaw_physics-based_2018,rodgers_broadband_2019,melgar_kinematic_2016}. Fault zones are the locus of intense deformation processes spanning a wide range of strain rates and contain valuable information on the long term history of these processes \citep{ben-zion_characterization_2003}; the geometry of a fault zone at a range of length scales, including any depth-dependent variations, can aid in reconstructing this history and constraining the physical processes involved \citep{norris_continental_2014,schulte-pelkum_tectonic_2020}.
        
        A fault zone's geometry is commonly assessed from a variety of sources. These include focal mechanisms determined with seismological methods \citep{lin_applying_2007,shelly2016new}, high-resolution seismicity catalogs \citep{chiaraluce_2016_2017,ross_abundant_2017}, various types of seismic imaging \citep{sato_earthquake_2005,fuis_subsurface_2017,lay_3d_2021,bangs_slow_2023}, geological data and mapped fault traces \citep{fletcher_assembly_2014}, and geodetic data \citep{lindsey_geodetic_2013}. These diverse information sources have their own uncertainties and sensitivities, making them complimentary when multiple sources are available; however it is not always straight forward to assimilate them. Several databases of fault geometry models have been produced with the goal of incorporating community consensus and providing established models with a documented provenance. These include faults at global scale \citep{bird_updated_2003,hayes_slab10_2012,hayes_slab2_2018} and also some regional scales \citep{plesch_community_2007,plesch_detailed_2020}.
        
        In this study, we aim to characterize the dip of fault zones in Southern California with high-resolution seismicity. We introduce a simple technique from the statistical field of spatial point processes that can measure fault zone dip independently from traditional methods, with the goal of augmenting the information available for constructing fault models. We first apply the method to four synthetic catalogs to demonstrate its suitability. We then apply the technique to eleven prominent seismicity regions across southern California to quantify the dip for different fault zone sections. These findings are compared with those of the SCEC Community Fault Model and other previous works in the area. We demonstrate that the method can reliably recover fault dip, including depth-dependent variations under some circumstances. Our primary scientific findings are that the San Jacinto and San Andreas fault zones appear to have significant northeasterly dips, whereas the Elsinore fault zone and Brawley Seismic Zone appear to be nearly vertical fault zones.
        
        \section{Methods}
        \subsection*{Preliminaries}
        Let $X \subset \mathbb{R}^D$ be a stochastic collection of points, i.e. a spatial point process \citep{daley_introduction_2003}. For a spatial domain $W \subset \mathbb{R}^D$, let $N(W)$ denote the number of points of $X$ that are contained within $W$. For those readers familiar with measure theory, $N(\cdot)$ is a counting measure on $W$. Since $X$ is a stochastic process, the mean number of points in $W$ is given by the so-called intensity measure,
        \begin{equation}
            \Lambda(W) = \mathbb{E} \left( N(W) \right),
        \end{equation}
        where $\mathbb{E}\left(\cdot\right)$ denotes an expected value. Let us also denote the volume of $W$ in $\mathbb{R}^D$ as $\left| W \right|$. Then, for a stationary point process, the quantity $\lambda = \Lambda(W) / \left| W \right|$ is independent of the choice of $W$. While $\Lambda(W)$ describes the expected number of points within a particular fixed volume, it does not describe spatial correlation of event density, i.e. knowing $\Lambda(W)$ does not tell you anything about $\Lambda(V)$ for some other disjoint $V \subset \mathbb{R}^D$.
        
        Instead, we need a different type of quantity to characterize the spatial correlation of points. For a typical point $u \in X$, one such choice is the $K$-function \citep{ripley_second-order_1976},
        \begin{equation}
            K(r) = \frac{1}{\lambda} \mathbb{E} \left( \text{number of neighbors within radius }r | X \text{ has a point at } u\right).
        \end{equation}
        The quantity $\lambda \, K(r)$ therefore quantifies the mean number of neighbors that any typical point will have within a sphere of radius $r$. The $K$-function is a cumulative function of $r$ and was first introduced to seismology by \citep{kagan_spatial_1980}, where it is often referred to as a correlation integral; most commonly the $K$-function has been used to infer the fractal distribution of a set of hypocenters by fitting a power law to an empirical estimator of the $K$-function. A useful property of $K(r)$ is that it describes how point patterns are arranged in space, independently of the choice of $W$. This is because $K(r)$ is a second-order quantity and is analogous to a covariance, whereas $\Lambda(W)$ is a first-order quantity and is analogous to an expected value.
        
        The function $K(r)$ has an inherent normalization property, which is seen by considering that for a Poisson process in 2D,
        \begin{equation}
            K_{pois}(r) = \pi r^2,
        \end{equation}
        i.e. $K_{pois}$ depends only on $r$ (and not on $\lambda$). This is important as it allows for $K_{pois}(r)$ to be used as a reference, and if $K(r) > K_{pois}(r)$, it is said that $X$ is clustered, since more of the points then locate within the sphere of radius $r$ than expected for the equivalent Poisson process. This is only possible because $K$ is conditional on a typical point existing at the center of the sphere.
        
        % As an example, consider the three different synthetic point patterns shown in Figure \ref{fig:pointpat}. Each pattern has $\lambda=100$ and has clustering present; the compactness of the clusters however increases progressively from A to B to C. Since $\lambda=100$ for all cases, it is not useful for discriminating between these cases, but, $K(r)$ is a second order quantity, and is clearly able to do so. Notice that as the degree of clustering increases between the patterns, this results in larger values of $K(r)$. It is this property that we rely on to quantify localization of seismicity.
        
        The $K$-function can be estimated using the following empirical formula,
        
        \begin{equation}
            \hat{K}(r) = \frac{\left| W \right|}{m(m-1)} \sum_i^m \sum_{j \neq i}^m \mathbf{1}\{d_{ij} \leq r\} e_{ij}.
            \label{eq:estimator}
        \end{equation}
        
        \noindent In this equation, $\mathbf{1}(\cdot)$ is the indicator function, $d_{ij}$ is the Euclidean distance between points $i$ and $j$, $e_{ij}$ is an edge correction factor, $m$ is the number of points in the observation window, and $\left| W \right|$ is the area (volume) of the observation window.
        
        \subsection*{The Cylindrical $K$-function}
        
        The $K$-function, as given above, is derived by assuming the point process is both stationary and isotropic, i.e. the likelihood of a point at $u$ given a point at $v$ depends only on the distance between them $r=\|u-v\|$. Seismicity, however exhibits strong spatial anisotropy at scales from local to global \citep{ross_geometrical_2022,nasirzadeh_new_2021,moller_geometric_2014,rubin_streaks_1999}. Seismicity lineations, i.e. collections of hypocenters that align in the form of linear features, are commonly observed in the highest resolution catalogs \citep{gillard_highly_1996,shearer_parallel_2002}. Sometimes, hypocenters align in the form of planar or surficial features \citep{ross_3d_2020,cox_injection-driven_2016}. Both linear and planar seismicity features are evidence of anisotropic point patterns since the likelihood of a point at a location $u$ given a point exists at $v$ depends on not just the spatial separation between them, but also the orientation of the vector connecting them, i.e. $K = K(u-v)$. 
        
\begin{figure}[htb]
            \centering
            \includegraphics[width=0.75\textwidth]{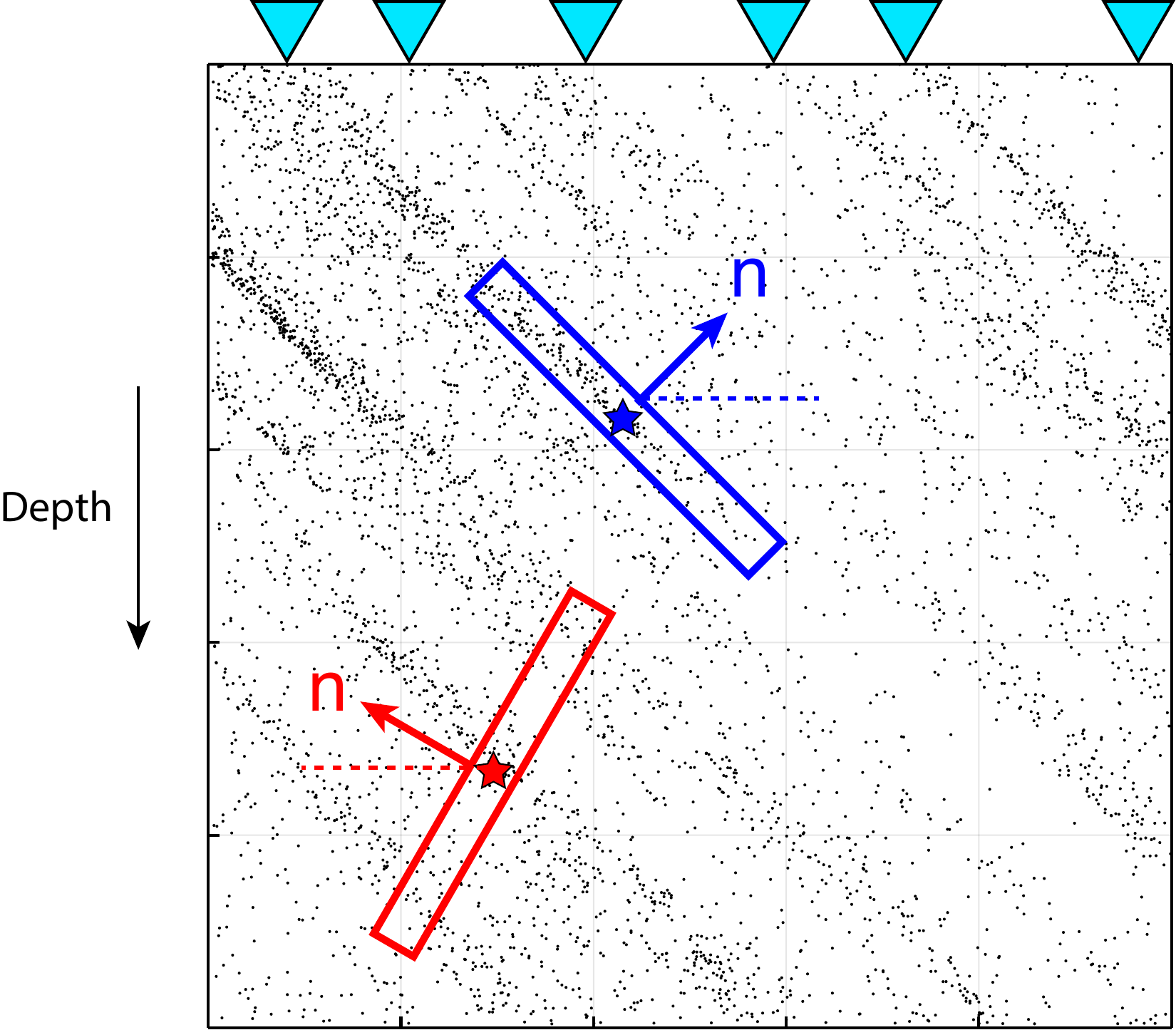}
            \caption{Illustration of method. A cylindrical K-function is computed by placing a disc with normal vector $\hat{n}$ centered on each event (stars). On average, the number of events contained in the disc is highest when the disc is aligned with seismicity lineations (blue box), resulting in a large value of $K$. Similarly, $K$ is low when poorly aligned with seismicity lineations (red box). The best dip estimate is equal to the dip of $\hat{n}$ for which $K$ is maximized. The method can detect dipping fabric even in distributed seismicity, such as in the cartoon, if a persistent orientation is present.}
            \label{fig:cartoon}
        \end{figure}

        Within the spatial statistics literature, there has been interest in detection and characterization of anisotropy in point processes \citep{moller_geometric_2014,moller_cylindrical_2016,safavimanesh_comparison_2016,nasirzadeh_new_2021}. One important development has been the cylindrical $K$-function \citep{moller_cylindrical_2016}, in which a cylinder is used in place of a sphere to characterize anisotropy that is effectively columnar. A cartoon example of this approach is shown in Figure \ref{fig:cartoon}, in which a cross section of seismicity is depicted. Here, the seismicity exhibits a dipping fabric that is orthogonal to the vector $\hat{n}$. When a cylinder defined by this normal vector is used (e.g. blue cylinder), the value of $K$ is maximized, as the cylinder on average will enclose more points than a cylinder aligned with any other orientation (e.g. red cylinder). By computing $K$ over all azimuths and polar angles, it is possible to detect anisotropy and quantify its orientation.
        
        For a unit vector $n = [\cos\varphi \sin\theta, \sin\varphi \sin\theta, \cos\theta]$, let $C_n (r, t)$ denote a cylinder with radius $r$, height $2t$, and normal vector $n$. For an observed set of points, $\{x_1, ..., x_m\}$, the cylindrical $K$-function \citep{moller_cylindrical_2016} is then computed as,
        \begin{equation}
            K_{cyl} (r, t, \theta, \varphi) = \frac{1}{\lambda^2} \sum_i^m \sum_{j \neq i}^m \mathbf{1}\{x_j - x_i \in C_n \} e_{ij},
            \label{eq:estimator_cyl}
        \end{equation}
        where the condition $x_j - x_i \in C_n$ is true if the vector separating $x_j$ and $x_i$ locates inside $C_n$, and $e_{ij}$ is an edge correction factor. In this study, we use the translation-based edge correction, a routine choice in point processes in which the window $W$ is translated by the vector $x_j - x_i$ and the amount of overlap between the translated window and the original window is computed,
        \begin{equation}
            e_{ij} = \frac{\left| W \right|}{\left| W \cap (W + x_j - x_i) \right|}
        \end{equation}
        
        We propose $K_{cyl}$ as a method to infer the dip of fault zones from seismicity, even when weakly localized as in Fig. \ref{fig:cartoon}, due to these aforementioned properties. While \citet{moller_cylindrical_2016} focused on detecting columnar structures with $K_{cyl}$ by using highly elongated cylinders (i.e. $r < t$), it can also be used to detect coherent surface-like structures in seismicity if the diameter of the cylinder is longer than its height (i.e. it is more aptly described as a disc, as in Fig. \ref{fig:cartoon}). This disc-based formulation is the one we use in this study.

        \subsection*{Demonstration with Synthetic Catalogs}
        
        We begin with four synthetically generated seismicity catalogs to demonstrate the method and provide additional insights into its usage. Furthermore, we use this opportunity to walk through the novel summary diagram used to visualize the results in this study.
        
        \begin{figure}[htb]
            \centering
            \includegraphics[width=\textwidth]{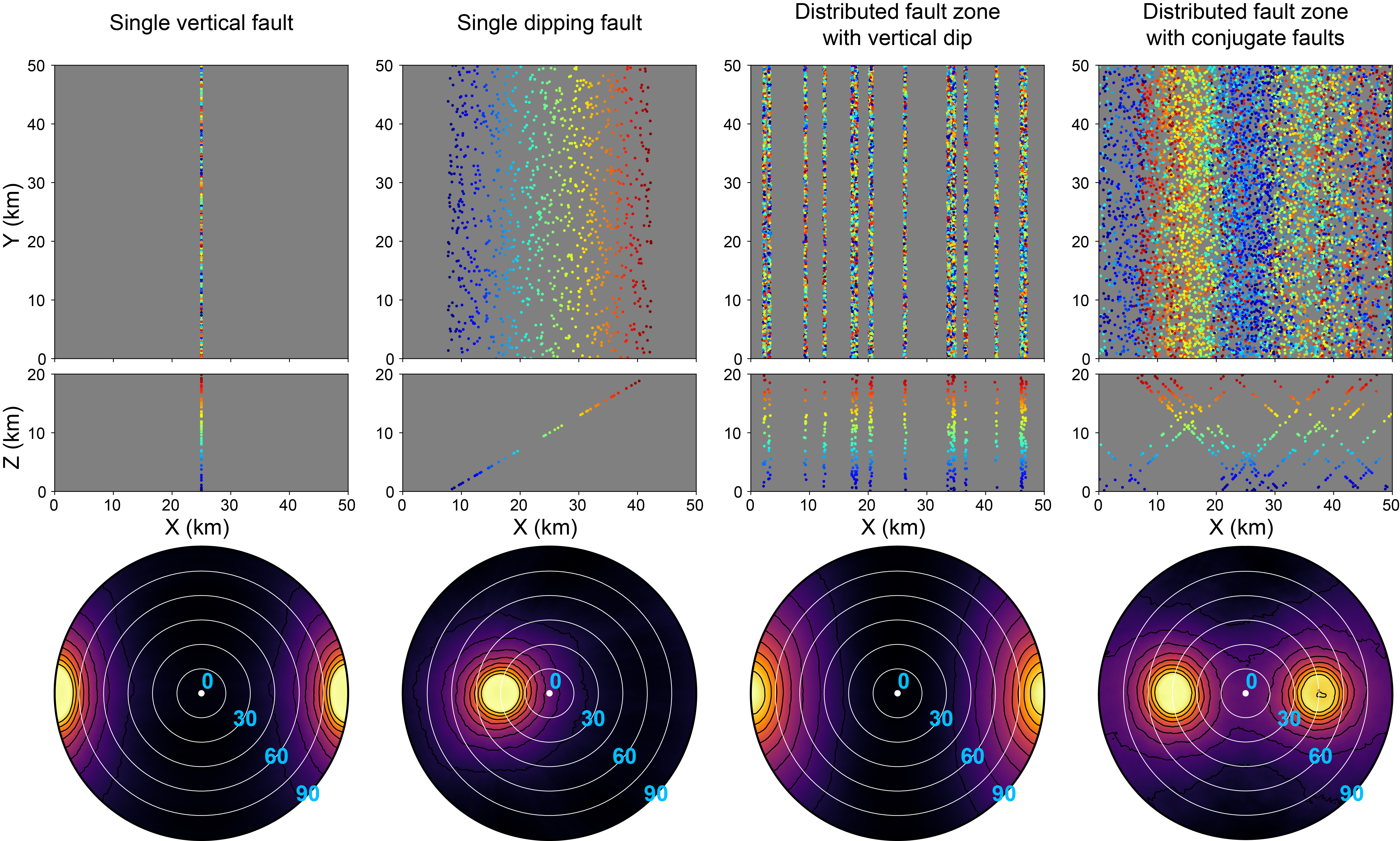}
            \caption{Method Demonstration with synthetic catalogs. Each column is a different seismicity catalog (described in main text). Events are colored by depth to enhance visibility. Upper row: map view of seismicity. Middle row: East-west cross section with seismicity projected onto it. For plotting purposes, seismicity is shown thinned by 95\%. Bottom row: Stereographic projection of $K_{cyl}$ for each catalog. Warmer colors indicate more intense clustering along a given fault normal azimuth and dip.}
            \label{fig:experiments}
        \end{figure}
        
        \noindent\textbf{Case 1: A single vertical planar fault}. We randomly generate 1000 hypocenters drawn from a uniform distribution on a planar N-S trending vertical fault with a length of 50 km and seismogenic thickness of 20 km. We set $r=0.1$km, $t=1.0$km, and compute $K_{cyl}$ on a grid with 2$^\circ$ spacing using equation \ref{eq:estimator_cyl}. Figure \ref{fig:experiments} shows the seismicity in both map view and cross-section. It also shows $K_{cyl}$ for this catalog in an upper-hemisphere stereographic projection, where the polar angle $\theta$ of the fault normal vector is given on the radial axis and the angle $\varphi$ is given as the traditional azimuthal angle for such a diagram. Here, $K_{cyl}$ correctly attains maxima at both $\varphi=90^\circ$ and $\varphi=180^\circ$, reflecting the symmetry of this particular dataset. The correct dip is also attained with little ambiguity.\\
        
        \noindent\textbf{Case 2: A single dipping planar fault}. We randomly generate 1000 hypocenters drawn from a uniform distribution on a N-S striking 30$^\circ$ dipping planar fault with a length of 50 km and seismogenic thickness of 20 km. As with the previous example, $K_{cyl}$ correctly recovers both the fault normal azimuth and the dip of the fault. Note that only one mode is present now in the $K_{cyl}$ plot, as the break in symmetry leads to the other mode occurring in the lower hemisphere, and thus not in the plot. \\
        
        \noindent\textbf{Case 3: Distributed fault zone with vertical dip}. We simulate seismicity occurring within a distributed fault zone having a vertical dip. Following the work of \citet{moller_cylindrical_2016}, we choose 20 random vertical faults (with dimensions 50 km $\times$ 20 km) that strike north-south. For each fault, we generate 500 random hypocenters that are then displaced randomly in the fault normal direction with Gaussian noise of 100 m to add complexity. The realization of this Poisson plane cluster process that we use is shown in Figure \ref{fig:experiments}. $K_{cyl}$ correctly identifies the same overall pattern as seen for the single planar vertical fault case, as there is just a single dominant orientation for the anisotropy even though 20 faults are present in the catalog. This demonstrates the potential for measuring fault dip even when the seismicity and fault zone is highly distributed, provided that the anisotropy is persistent across much of the seismicity. \\

        \noindent\textbf{Case 4: Distributed fault zone with conjugate faults}. We simulate seismicity occurring within a distributed fault zone having conjugate faults with dips of around 45$^\circ$. The dip is randomly perturbed so that not all angles are identically 45$^\circ$. We create 20 faults that strike north-south, with half dipping to the west and half dipping to the east. For each fault, we randomly locate 500 hypocenters within it. The hypocenters are then displaced randomly in the fault normal direction with Gaussian noise of 100 m to add complexity. The resulting catalog is shown in Figure \ref{fig:experiments}. $K_{cyl}$ correctly indicates two orthogonally dipping faults with the same strike. This demonstrates the potential for measuring multiple fault dip angles, when present. \\
       
        \subsection*{Application to Southern California Seismicity}
\begin{figure}[htb]
            \centering
            \includegraphics[width=0.95\textwidth]{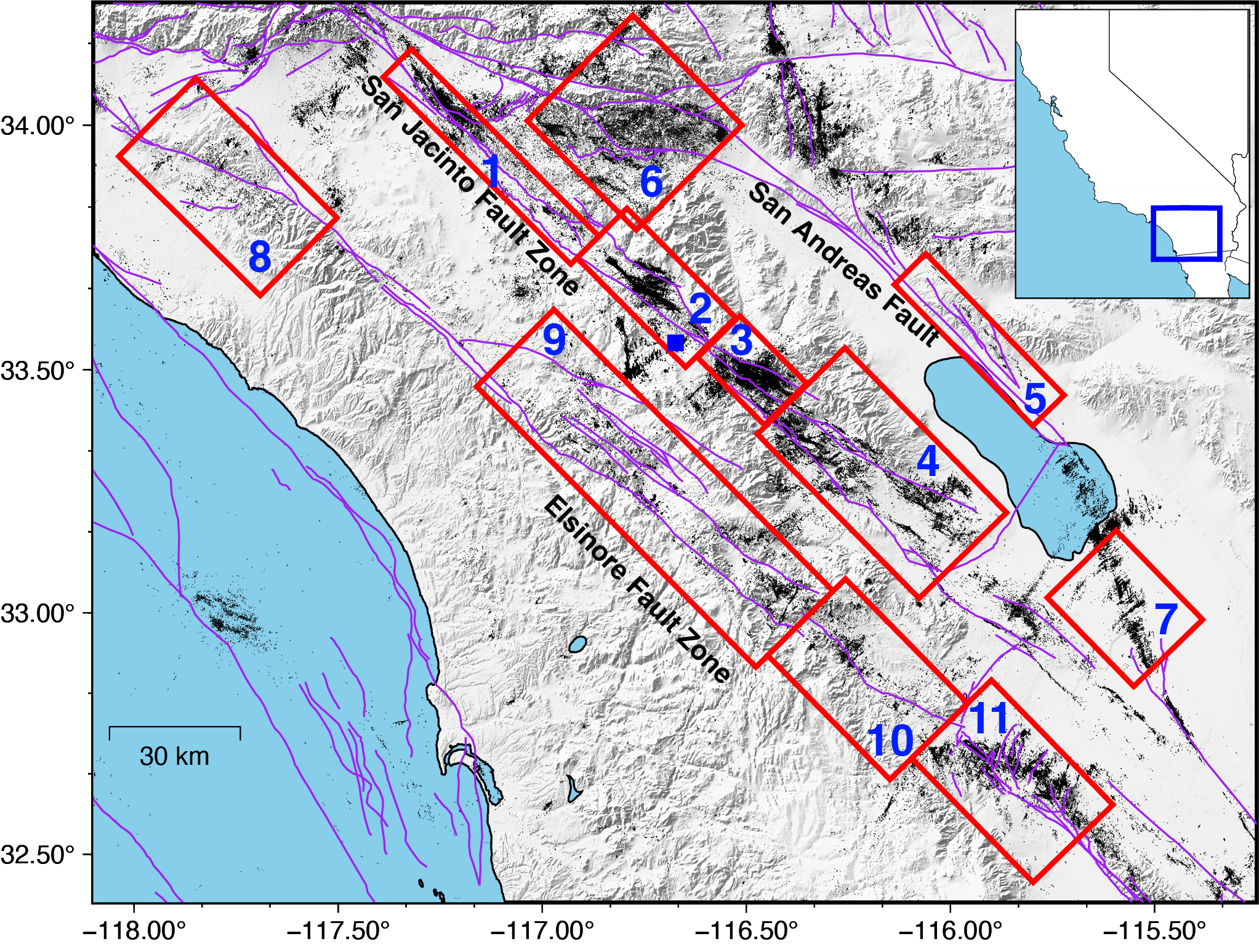}
            \caption{Map of seismicity in Southern California. Black dots indicate relocated epicenters. Red lines denote focus areas with numbers matching region names provided in Table \ref{table:regions}. Blue square indicates the town of Anza, California.}
            \label{fig:map_socal}
        \end{figure}

        We now shift our focus to using $K_{cyl}$ to quantify the dip for fault zones in Southern California. We use a high-resolution relocated seismicity catalog that covers the entirety of southern California and the northern part of Baja California for the period 1981-2019. The catalog used is based on the methodology of \citet{hauksson_waveform_2012} and has been updated for recent years (Fig. \ref{fig:map_socal}). It contains contains 679,495 earthquakes that have been relocated with waveform cross-correlation, which form the highest quality subset. We focus only on the relocated events in this study. The catalog is publicly available from the Southern California Earthquake Data Center \citep{southern_california_seismic_network_southern_2013}. We use only the hypocenters and magnitudes for these catalogs. 
        
        We also considered using the the Quake Template Matching (QTM) catalog for southern California \citep{ross_searching_2019}, which contains 10 times more events but spans only the period 2008-2017. Ultimately, we opted for the \citet{hauksson_waveform_2012} catalog because it is much longer in duration and the hypocenters are generally more precise; the many extra smaller events detected in the QTM catalog have fewer phase picks available and lead to an overall slight degradation in location accuracy as compared with the \citet{hauksson_waveform_2012} events, which is less desirable for this study.
        
        For our analyses, we subset the catalog into 11 non-overlapping fault zone sections. They are denoted by red boxes in Fig. \ref{fig:map_socal} and described in more detail in Table \ref{table:regions}; the number of earthquakes within each region is also given. These regions were chosen based on a variety of factors, including scientific or hazard importance, longstanding fault segment demarcation by the community, an abundance of seismicity, or clear geometrical boundaries. The list contains four sections of the San Jacinto Fault Zone, two sections of the San Andreas Fault Zone, four sections of the Elsinore Fault Zone, and the Brawley Seismic Zone. For all but one of the regions, there are thousands of earthquakes available, which is important to ensure the statistical estimators are robust.  
        
        For each region, we compute $K_{cyl}$ using the horizontal coordinates as defined in Fig. \ref{fig:map_socal} and using the depth range [0,22] km. We then use equation \ref{eq:estimator_cyl} to compute for three sets of parameters, ($t$, $r$) = (50 m, 500 m), (100 m, 1000 m), and (200 m, 2000 m). We compute $K_{cyl}$ for $\theta \in [0, \pi]$ and $\varphi \in [0, 2\pi]$, i.e. the whole range possible, as we aim to estimate the dip of each fault zone without any prior knowledge. This framework also provides a means to perform hypothesis testing if several candidate scenarios for the dip are believed to be possible (which is covered in more detail in the discussion). The domains for $\theta$ and $\varphi$ are discretized with spacing of $1^{\circ}$; this choice is mainly a balance between having sufficiently fine spatial resolution and computational efficiency, since the results are largely insensitive to them. Given $K_{cyl}$, the best estimate of the fault normal vector is defined by the values of $\theta$ and $\varphi$ for which $K_{cyl}$ is maximized (as in Fig. \ref{fig:cartoon}). The best fault zone dip estimate is then $\delta=\pi-\theta$.
        
        \begin{table}[htb]
        \centering
        \caption{Description of the focus areas in Southern California}
        \begin{tabular}{@{}clc@{}}
        \toprule
        Region \# & Region Name                                & Number of Events \\ \midrule
        1         & San Jacinto Fault Zone (Claremont)         & 14,340                 \\
        2         & San Jacinto Fault Zone (Hot Springs)       & 24,066                 \\
        3         & San Jacinto Fault Zone (Trifurcation Area) & 29,914              \\
        4         & San Jacinto Fault Zone (Borrego Mountain)  & 24,662                 \\
        5         & Southern San Andreas                       & 723                 \\
        6         & San Gorgonio Pass                          & 23,614                 \\
        7         & Brawley Seismic Zone                       & 9,402                  \\
        8         & Elsinore Fault Zone (Whittier)             & 3,396                 \\
        9         & Elsinore Fault Zone (Julian)               & 17,644                 \\
        10        & Elsinore Fault Zone (Coyote Mountain)      & 6,864                 \\
        11        & Elsinore Fault Zone (Yuha)                 &  21,939                \\ \bottomrule
        \end{tabular}
        \label{table:regions}
        \end{table}

       \subsection*{Dip Uncertainty Estimates}
        The polar diagrams for $K_{cyl}$ are useful for visual examination of the results and identifying the most likely dip angle(s), but do not communicate the uncertainty associated with these measurements. To obtain uncertainty estimates, we use a bootstrapping approach designed for spatial resampling of these empirical estimators \citep{loh_valid_2008}. We use this method to resample local $K_{cyl}$ functions with replacement, compute an average $K_{cyl}$ function for each bootstrap sample, measure $\delta=\pi-\theta$ corresponding to the peak of $K_{cyl}$, and repeat this process 1000 times. The ensemble of $\delta$ values resulting from the bootstrap procedure provides an estimate of the uncertainty.

        \subsection*{Parameter Selection and Resolution}
\begin{figure}[htb]
            \centering
            \includegraphics[width=0.65\textwidth]{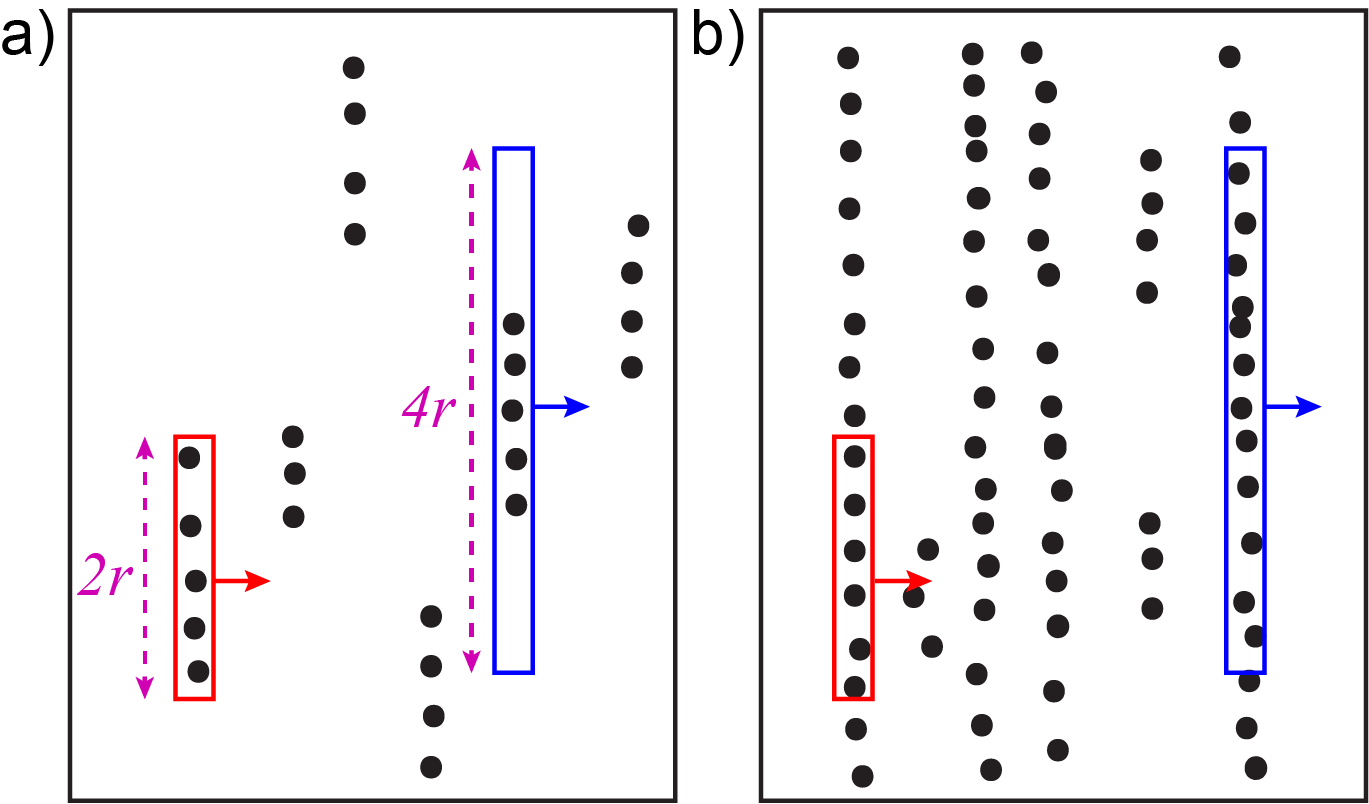}
            \caption{Cartoon illustrating the spatial resolution of the method. In a), the point pattern has an effective length scale of less than $2r$, and the pattern can be resolved by $K_cyl$ to $\leq2r$. In b), the pattern has an effective length scale generally larger than $4r$, but with the two discs shown, the pattern can only be resolved to $\leq4r$.}
            \label{fig:cartoon2}
        \end{figure}

        The two parameters $t$ and $r$ control the resolution of the method and here we give some additional insight and guidance around their usage. Generally speaking, it will be unknown beforehand what length scales are useful for measuring the dip. Thus, it is desirable to to compute $K_{cyl}$ for a range of values. Figure \ref{fig:cartoon2} shows two schematic scenarios and the potential for resolving faults with the method. In Fig \ref{fig:cartoon2}, a red disc of radius $r$ and a blue disc of radius $2r$ are shown, with $t \ll r$ for both. In (a), the seismicity pattern has structure with an effective length scale of about $2r$. For this case, both the red and blue discs can resolve this anisotropy since the length scale is less than or equal to the diameter of the disc. Thus, the diameter of the disc is effectively an upper bound to the length scale of the anisotropy. In (b), the seismicity pattern exhibits a length scale comparable to the whole window. In this case, both the red and blue discs can resolve the anisotropy, however since both discs have a diameter smaller than the length scale of the seismicity, they are unable to provide information about larger length scales.
        
        If the true hypocenter configuration exhibits planar anisotropy, then making the disc thickness $t$ as small as possible will increase sensitivity for detecting anisotropy. However, the lower limit for whether $t$ will be useful is closely related to the location errors in the respective direction. Thus, we recommend initially setting the value of $t$ to be comparable to the estimated relative location error of most events.

        Practically speaking, there will be limits to the value of $r$ that can be used. The largest values of $r$ used should depend on the dimensions of the spatial window, $W$; in particular, $K_{cyl}$ will become unreliable as $2r$ approaches values of roughly 1/4 the shortest spatial dimension of $W$. This is true despite the use of an edge correction factor, as there will be little usable signal left to correct at these scales, similar to amplifying noise in seismic deconvolution. At the same time, $r$ should still be much larger than $t$, in order to have sufficient sensitivity in detecting anisotropy. As the aspect ratio $r/t$ approaches 1:1, $K_{cyl}$ becomes effectively unable to identify anisotropy. Additionally, $r$ should be large enough that enough events locate within the discs to constrain $K_{cyl}$ to a desirable level (preferably as measured from the aforementioned bootstrap procedure).

        For this study, we use a single fixed aspect ratio of $r/t = 10$, in part to simplify the process of choosing these parameters. This allows for the same level of statistical power in resolving anisotropy, while still allowing the spatial resolution to vary. Larger aspect ratios may lead to similar results for the regions in which there are plentiful events. Given the variably-sized regions in Figure \ref{fig:map_socal}, the smallest regions will have the lowest maximum values of $r$. In an effort to ensure uniformity across the regions, we chose a maximum value of $r=2$km, which results in a value of $t=200$m. We then decreased $r$ by powers of 2, which results in $(r,t)=(1000m,100m), (500m, 50m)$. The latter of these parameter pairs is essentially the lower limit of what is possible, and still have enough points to resolve $K_{cyl}$.

\begin{figure}[htb]
            \centering
            \includegraphics[width=\textwidth]{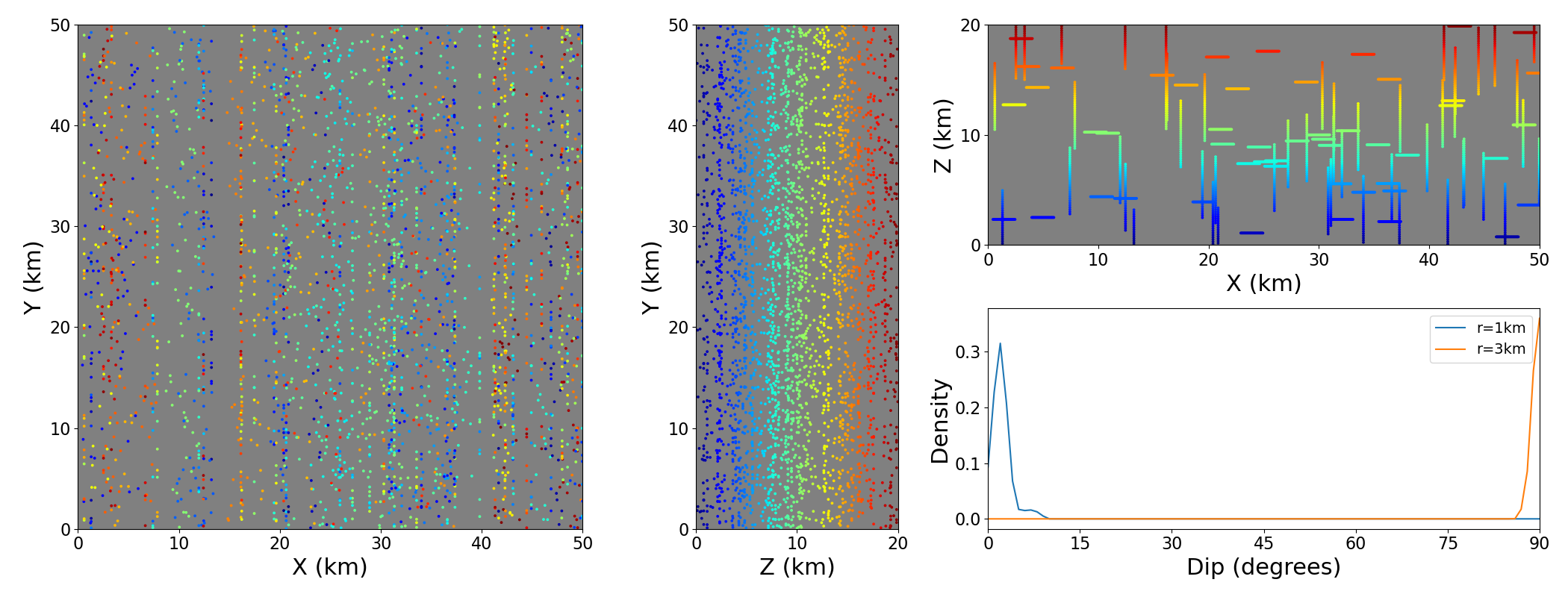}
            \caption{Synthetic catalog demonstration of two fault dip orientations at different length scales (1km and 3km, respectively). Events are colored by depth to enhance visibility. Lower right panel shows bootstrap recovery results for $K_{cyl}$ at two different length scales.}
            \label{fig:multiscale}
        \end{figure}

        Since $K_{cyl}$ is a cumulative function of $r$ and $t$, there may be questions relating to the ability for it to resolve different dip values if present at strictly different length scales. Indeed using such cumulative descriptive metrics is not ideal for this case; a more suitable quantity for this scenario may be the anisotropic pair correlation function, \citep{moller_geometric_2014,ross_geometrical_2022}. However, $K_{cyl}$ can still be of some use, depending on the circumstances. To show this, we create a simple synthetic catalog consisting of vertical and horizontal faults having the same strike, as in Fig. \ref{fig:multiscale}. Here, the vertically dipping faults have an effective length scale of 3 km whereas the horizontal faults have a length scale of 1 km. We compute $K_{cyl}$ for this dataset using $t=0.25$km and two values of $r$, $r=1$km, $r=3$km. A bootstrap analysis is used to show the dip uncertainty estimates for each value of $r$. Indeed both faults are reliably recovered.

        \section{Results}
        In this section we summarize the main findings for each region and evaluate them in the context of information available from other sources and methods. For southern California, the most comprehensive resource available documenting fault zones and their geometry is the Community Fault Model (CFM) produced by the Southern California Earthquake Center (SCEC) \citep{plesch_community_2007}. This database has been assembled by the SCEC community from a multitude of data sources including focal mechanisms, seismicity, seismic data, geology, and geodetic deformation. The CFM has comprehensive coverage across southern California, and we use version 5.3 \citep{plesch2020community} as a baseline for evaluating our results. In addition, we compare our results to those of other studies whenever available, on a case-by-case basis. Next, we walk through the results for each fault zone.
        
        \subsection*{San Jacinto fault zone}
        The San Jacinto fault zone (SJFZ) is a major strike-slip system in the southern California plate boundary area that branches off from the San Andreas in the Cajon Pass and extends southeast to the Imperial Valley. The SJFZ has multiple primary strands and several major stepovers \citep{sharp_san_1967}. Northwest of the town of Anza, the Clark fault is believed to be the main seismogenic structure of the SJFZ \citep{share_internal_2017}, whereas just southeast of Anza, the Coyote Creek fault branches off of the Clark fault and takes over as the primary fault \citep{qiu_internal_2017}. The seismicity in the SJFZ tends to exhibit weak spatial clustering but strong geometric anisotropy \citep{ross_geometrical_2022}. The SJFZ exhibits considerable variation in the seismogenic depth along-strike that is attributed to variations in heat flow \citep{doser_depth_1986}, with depths approaching 20 km at the northwest end in the Cajon Pass, to roughly 10 km near the Salton Trough. While historically considered to be a nearly vertical fault zone, more recent works have concluded that the main structures in the central SJFZ are dipping to the northeast, particularly at depth \citep{plesch2020community,ross_abundant_2017,schulte-pelkum_tectonic_2020}. \citet{schulte-pelkum_tectonic_2020} conclude that most of the central SJFZ is dipping NE in the range $\sim65^\circ-80^{\circ}$. 
        
\begin{figure}[h]
            \centering
            \includegraphics[width=0.9\textwidth]{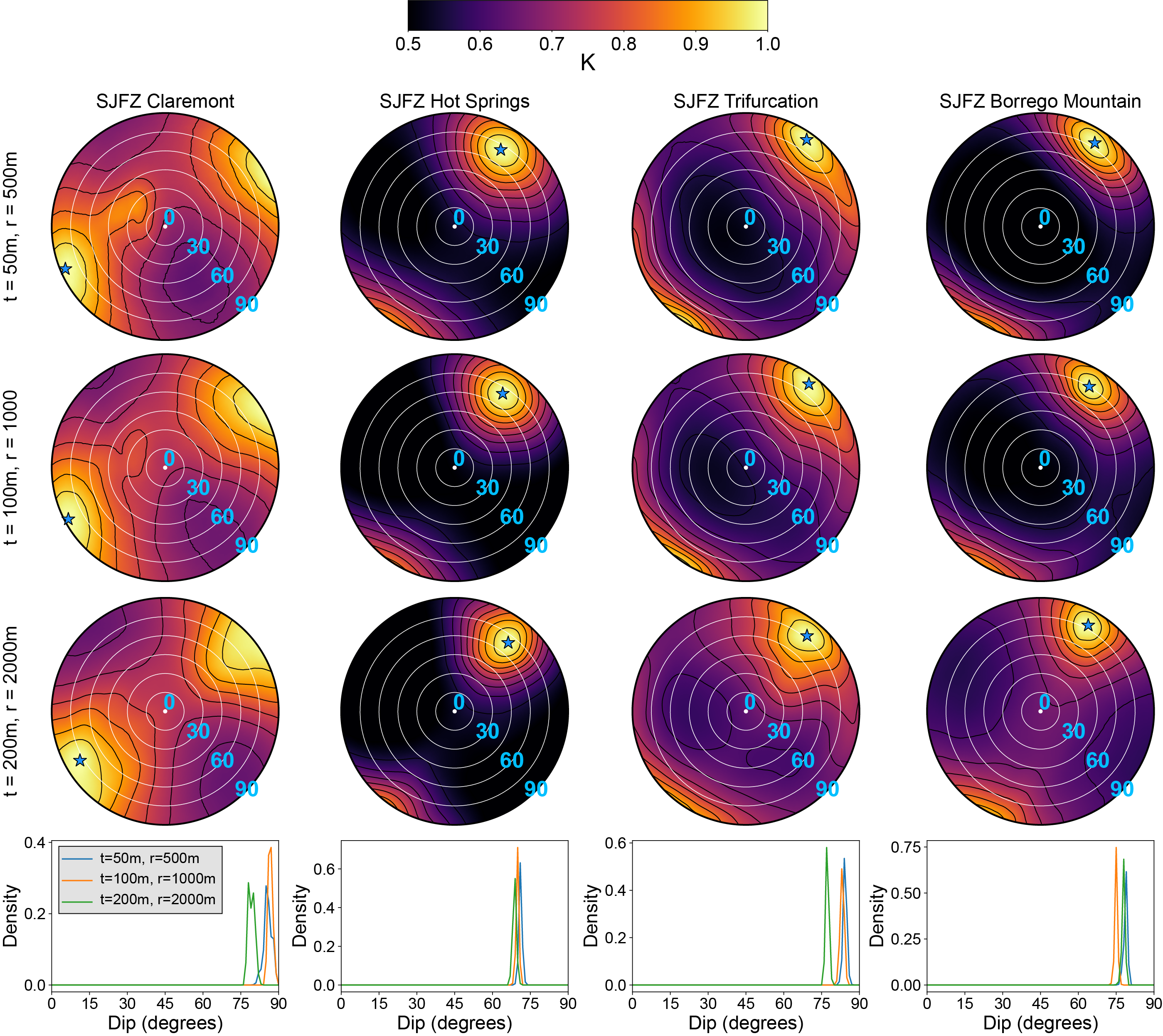}
            \caption{Cylindrical K-functions for the San Jacinto Fault Zone and dip estimates. Density functions for each region (bottom row) are bootstrap distributions for best dip estimate. These areas trend from northwest to southeast. The Claremont section is nearly vertical on average, whereas the other three sections dip moderately to the northeast.}
            \label{fig:sjfz}
        \end{figure}

        We analyze four key seismicity regions of the SJFZ in Fig. \ref{fig:sjfz} (see also Table \ref{table:regions}) with cylindrical K-functions: Claremont, Hot Springs, Trifurcation area, and Borrego Mountain. The results in Fig. \ref{fig:sjfz} are computed over the entire [0,22] km depth range, and should therefore be interpreted as average values; however it should be noted that for the SJFZ, seismicity generally does not occur above 5 km or so \citep{hauksson_applying_2019}, and thus the results largely reflect the deeper part of the fault zone. Each row uses a different combination of $(t, r)$. We notice from the diagrams that in each case, the largest value of $K_{cyl}$ indicates a fault normal azimuth in the range of $29^\circ-64^\circ$. In fact, except for the Claremont section, the SJFZ regions have a consistent estimate of the fault normal azimuth in the range $29^\circ-39^{\circ}$. The radius of the polar plot indicates the dip of the normal vector, and can be used to estimate the average dip of the fault zone; the bootstrap histograms in the bottom row of Figure \ref{fig:sjfz} show the estimated dips and their uncertainties. In the Hot Springs section, $\delta=68^\circ-72^{\circ}$ NE, the Trifurcation area estimates are $\delta=77^\circ-84^{\circ}$ NE, and the Borrego Mountain estimates are $\delta=75-79^{\circ}$ NE. The SCEC CFM has most of these faults listed as subvertical NE dipping faults, with the Hot Springs, Trifurcation, and Borrego Mountain dip values given as $\delta=82^{\circ}$ NE, $\delta=88^\circ-89^{\circ}$ NE, and $\delta=88^\circ-89^{\circ}$ NE, respectively. However, our results for the Claremont section indicate the opposite sense of dip, with $\delta$ estimated to be $78^\circ-88^{\circ}$ SW; this is in fact close to the CFM results, which has $\delta=84^{\circ}$ SW. The results in this figure have effective length scales of 1, 2, and 4 km, and since there is little variation in the dip for these different parameters, they indicate that the dip estimates are robust at these scales. The results do not imply anything about dip at larger scales.
        
\begin{figure}[htb]
            \centering
            \includegraphics[width=0.95\textwidth]{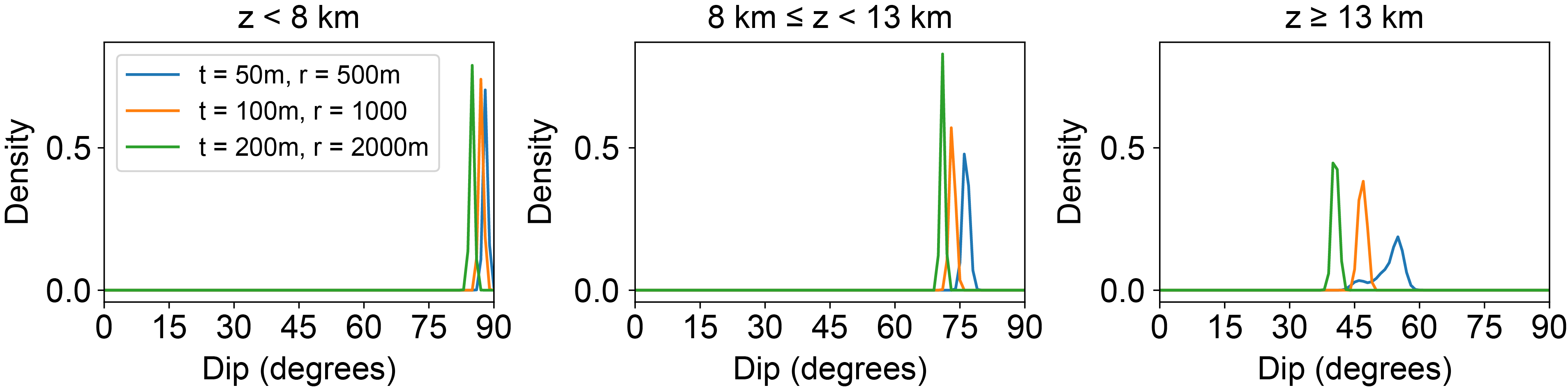}
            \caption{Estimating the depth dependence of $\delta$ for the SJFZ Trifurcation Area. This section of the fault zone exhibits evidence of listric strike-slip behavior. Left, middle, and right panels use 5584, 16862, and 7466 events, respectively.}
            \label{fig:trif}
        \end{figure}

        The abundance of seismicity in the central SJFZ allows us to further quantify the dip in depth slices to look for possible depth-dependent variations. \citet{ross_abundant_2017} argued the SJFZ trifurcation area exhibits listric-type behavior based on combined examination of relocated seismicity, focal mechanisms, and mapped surface fault traces. \citet{ross_abundant_2017} concluded that the SJFZ is nearly vertical in the upper 10 km and dipping $70^\circ$ NE below this. Here, we independently investigate this idea with $K_{cyl}$ by splitting the seismicity into three depth bins: 0-8 km, 8-13 km, and $>$13 km, containing 5584, 16862, and 7466 events, respectively. Figure \ref{fig:trif} shows $K_{cyl}$ for the three depth bins. The best estimates of $\delta$ are $88^\circ$ NE, $76^\circ$ NE, and $53^\circ$ NE, respectively, which indeed suggest that the fault zone is listric in this area, consistent with the conclusions of \citet{ross_abundant_2017}. For cross sections of the seismicity in this area, the reader is recommended to see Figure 7 of \citet{schulte-pelkum_tectonic_2020} or Figure 2 of \citet{ross_abundant_2017}. 

        \subsection*{San Andreas Fault Zone}
        The portion of the San Andreas Fault Zone (SAFZ) from the Cajon Pass to its terminus at Bombay Beach is just one of the three major sub-parallel strike-slip systems in southern California. There are important questions about its geometry along this part of the plate boundary and it has been the subject of extensive analysis \citep{fuis_new_2012,fuis_subsurface_2017,lindsey_geodetic_2013,fattaruso_sensitivity_2014,schulte-pelkum_tectonic_2020}, much of which has focused on whether the main seismogenic fault is vertical or dipping northeast, a question that is of prime importance for earthquake rupture simulations as it will affect both the magnitude of potential earthquakes and also the shaking pattern \citep{graves_broadband_2008,graves_shakeout_2011}.
        
        The San Gorgonio Pass (SGP) region of the SAFZ is concentrated around the San Bernardino Mountains. The seismicity here is weakly clustered spatially \citep{ross_geometrical_2022} and extends down to a depth of $\sim20$~km, the effective lower limit for seismicity in southern California \citep{hauksson_waveform_2012}. The slip rate in this area is about 24 mm/year and there are several major strands: the Mission Creek, Banning, and Garnet Hill faults \citep{gold_holocene_2015,fuis_subsurface_2017,blisniuk_revised_2021}. There are also numerous minor strands that may not extend to the surface \citep{fuis_subsurface_2017,schulte-pelkum_tectonic_2020}. Since the start of the instrumental era of seismology in southern California, two significant earthquakes occurred in this area, 1948 $M_L$ 6.5 Desert Hot Springs \citep{richter_desert_1958,nicholson_seismic_1996} and 1986 $M_w$ 6.0 North Palm Springs \citep{jones_north_1986,nicholson_seismic_1996,mori_source_1990}.
        
\begin{figure}[htb]
            \centering
            \includegraphics[width=0.80\textwidth]{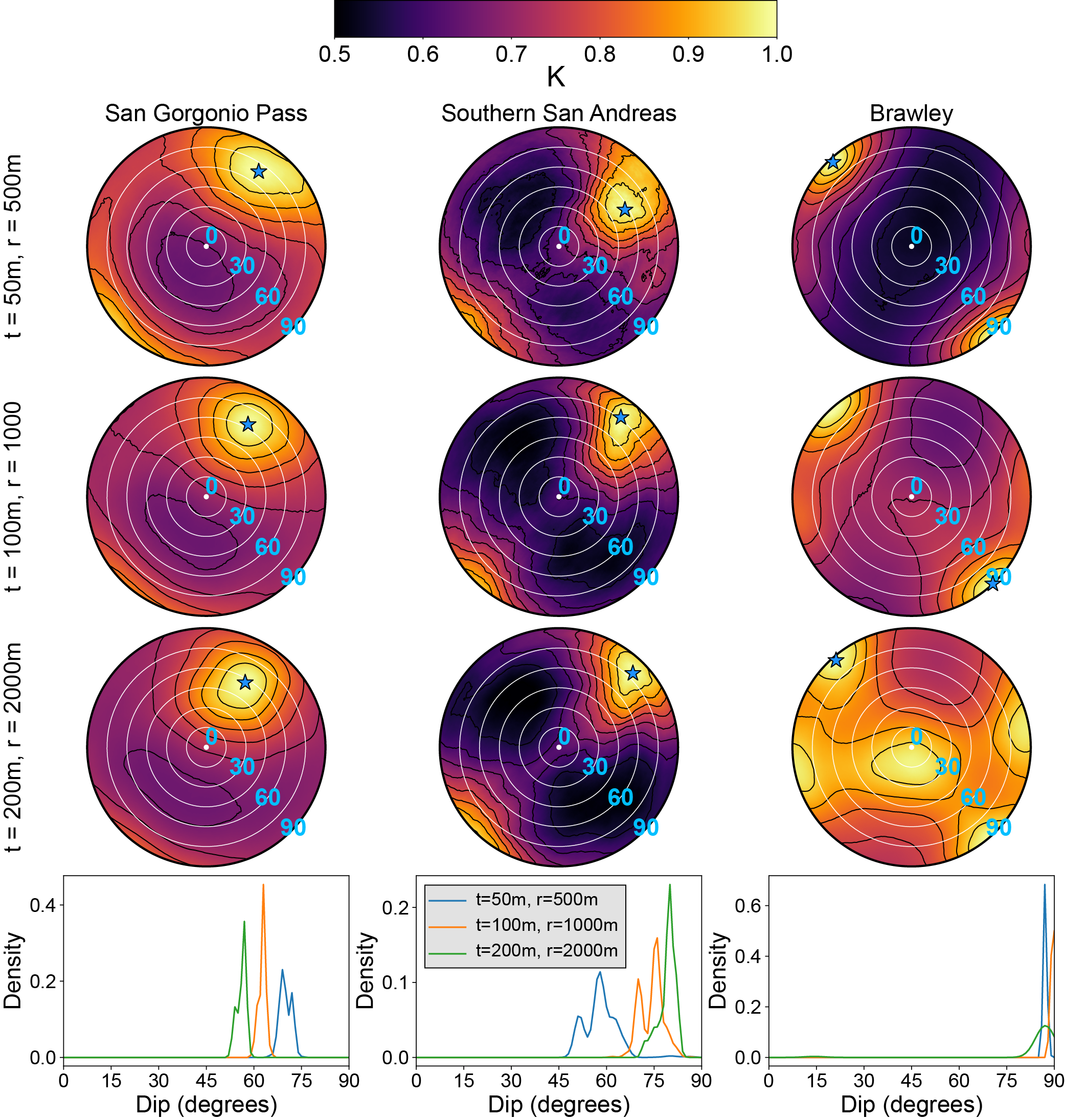}
            \caption{Cylindrical K-functions for the San Andreas Fault Zone and Brawley Seismic Zone. SAFZ seismicity dips to the northeast.}
            \label{fig:ssaf}
        \end{figure}

        Figure \ref{fig:ssaf} shows $K_{cyl}$ results for the SGP region. The estimates of $\varphi$ and $\delta$ indicate a NE dipping fault zone, with $\delta$ in the range $54^\circ-70^\circ$, depending on the scale of the cylindrical elements used. More specifically, we find that $\delta$ decreases as the length scale is increased, which suggests that the larger (older) structures in this fault zone are oriented more horizontally, whereas the younger (smaller) structures are slightly more vertical. For comparison, the CFM \citep{plesch2020community} has the Banning Fault dipping $72^{\circ}$ NE and the Mission Creek Fault dipping $82^{\circ}$ NE. \citet{fuis_subsurface_2017} identify seismic reflectors in this area that are dipping in the range $\sim55^\circ-65^\circ$ NE, with some more steeply dipping structures too. The 1948 $M_L$ 6.5 and 1986 $M_w$ 6.0 mainshocks in this area have focal mechanism dips of about $45^{\circ}$ \citep{jones_north_1986,nicholson_seismic_1996}. Our results reflect average values of fault zone dip over the entire SGP region, which includes many smaller structures in between the Banning and Mission Creek faults.
        
        Southeast of the SGP is the Coachella Valley section (Southern San Andreas) of the SAFZ. This portion runs from about Palm Springs to Bombay Beach, the southernmost terminus of the system. In this section also, there is debate over whether the fault zone is dipping \citep{fuis_subsurface_2017,lin_applying_2007,schulte-pelkum_tectonic_2020}. The SCEC CFM 5.3 has the Southern San Andreas fault as being pure vertical ($\delta=90^\circ$), whereas others including \citet{fuis_subsurface_2017,lindsey_geodetic_2013} conclude the SAFZ dips $\sim50^\circ-60^\circ$ NE. Our $K_{cyl}$ results for the Coachella section of the SAFZ are shown in Fig. \ref{fig:ssaf}. The method unambiguously identifies a NE dipping fault zone. At the smallest length scale examined, $r=50$ m, $t=500$ m, our best estimate of $\delta$ is just under $60^\circ$ NE. However, as the scale increases, so does $\delta$: for $r=100$ m, $t=1000$ m, $\delta=73^{\circ}$ NE, and for the largest scale, $r=200$ m, $t=2000$ m, our best estimate of $\delta$ is $80^{\circ}$.
        
        The trend of $\delta$ increasing with scale for the Southern San Andreas is opposite to what was observed for the SGP. We interpret these deviations between the smallest and largest scales to reflect down-dip curvature of the fault zone, with a listric type behavior that is more vertical in the upper $\sim8-10$ km and more horizontal below this. However it is important to remember that all scales exhibit clear evidence of a northeast dipping fault zone.
        
        \subsection*{Brawley Seismic Zone}
        The Brawley Seismic Zone is one of the more complex faulting regions in California, serving as the plate boundary transition between the SAFZ and the Imperial and Cierro Prieto faults in Baja California. The region is known for having considerable swarm activity \citep{hauksson_report_2013,hauksson_evolution_2017,hauksson_seismicity_2022}, conjugate/orthogonal faults \citep{thatcher_fault_1991,ross_geometrical_2022}, and prolific geothermal activity \citep{brodsky_anthropogenic_2013}.
        
        The SCEC CFM lists all of the major faults in the Brawley Seismic Zone as being vertical strike-slip. Our findings for this region are shown in Fig. \ref{fig:ssaf} and have dip estimates that are relatively consistent between the three different length scales. However, there are clear differences in the strike distribution between these scales; the $K_{cyl}$ identifies two clear modes in the strike distribution for $(t=200 m, r=2000 m)$ with roughly equal occurrence, separated by about 60$^\circ$ in azimuth. The conjugate faulting eventually disappears for $(t=50 m, r=500 m)$ and a NW-SE trending orientation is the only one visible. Thus, we can say quantitatively that the NW-SE structures are generally larger than 2 km in length. This orientation is the most closely aligned with the Southern San Andreas, and may reflect the current orientation that new damage and cracking is being produced for. This might imply that the NW-SE trending seismicity structures are relic structure from previous faulting that has not healed.
        
        \subsection*{Elsinore Fault Zone}
        
        The Elsinore Fault Zone (EFZ) is the youngest of the three major fault systems composing the southern California plate boundary. The EFZ also has the lowest slip rates of the three, being $\sim 5$mm/year \citep{magistrale_central_1996}. In the northwest, the EFZ emerges near the eastern end of the Los Angeles basin and extends southeast for roughly 200 km before becoming the Laguna Salada fault zone near the United States-Mexico border. EFZ seismicity is more scarce compared with some of the other regions we have examined, and so we examine here four sections that have sufficient events to perform a $K_{cyl}$ analysis. 
        
\begin{figure}[h]
            \centering
            \includegraphics[width=0.9\textwidth]{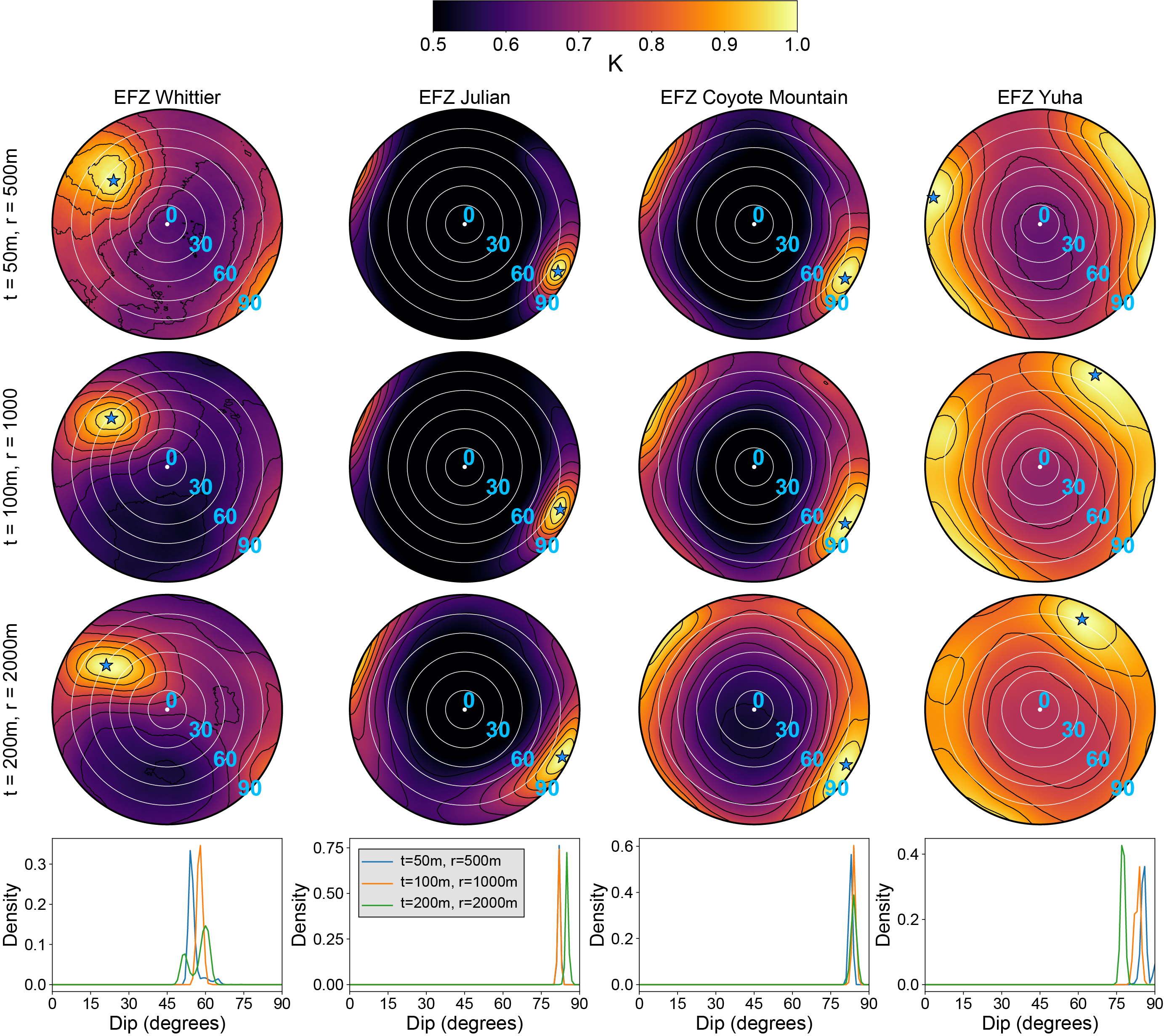}
            \caption{Cylindrical K-functions for the Elsinore Fault Zone. Regions in the first three columns exhibit prominent seismicity anisotropy that is orthogonal to the main strike of EFZ. Most of the EFZ seismicity has nearly vertical dip, except Whittier section. Yuha Desert section has conjugate seismicity with a high angle.}
            \label{fig:elsinore}
        \end{figure}

        The Whittier section of the EFZ is located in the eastern LA Basin and is viewed as a transition region from the compressional regime of the transverse ranges to the strike-slip regime of the Elsinore system \citep{hauksson_earthquakes_1990}. The Whittier fault branches off from the dominant trend of the EFZ at an angle of $\sim15^{\circ}$ and has a strike of about $300^{\circ}$. Beneath the Whittier fault is the Puente Hills blind thrust \citep{shaw_elusive_1999}. The Whittier fault is listed in the SCEC CFM as dipping to the northeast at $77^{\circ}$. Events in the area typically have focal mechanisms with considerable obliquity \citep{yang_evidence_2011}, with the largest in recent memory being the 2008 $M_w$ 5.4 Chino Hills earthquake \citep{hauksson_preliminary_2008}. There has been some discussion of the orientation of the structures here, with both nodal planes being considered as plausible. \citet{shao_rupture_2012} analyzed the kinematic rupture process of the Chino Hills earthquake and tested both nodal planes, concluding that the plane aligned with the Whittier fault was most likely. Figure \ref{fig:elsinore} shows our $K_{cyl}$ results for the Whittier, which indicates for all three scales a NW fault zone dipping $51^\circ-64^\circ$ and a strike of $34^\circ-40^\circ$. These values are close to the parameters of the "auxiliary plane" for most focal mechanisms in the area; for example the Chino Hills earthquake had an auxiliary plane with a strike of $42^{\circ}$ and a dip of $55^{\circ}$. Importantly, $K_{cyl}$ does not show any evidence of a second mode aligned with the Whittier fault. From this, we thus conclude that at least at the scale of 1-4 km, the active seismogenic structures in the area are a mixture of left-lateral and thrust slip that are not aligned with the Whittier fault. At larger scales, it is very possible that fault zone structures align with the Whittier fault and dip northeast as given in the CFM.
        
        The Julian and Coyote Mountain sections cover most of the central and southern EFZ. The major fault traces within these sections are relatively straight and trend southeast. Both sections are listed in the CFM as being nearly vertical (81-87$^{\circ}$) with a strike of around 305$^{\circ}$. For these sections, the peak $K_{cyl}$ value (Fig. \ref{fig:elsinore}) indicates a strike of $204^\circ-210^\circ$ and a dip of $82^\circ-86^\circ$; thus our results identify the orthogonal plane as being the dominant one visible in the seismicity at the scale of 1-4 km. This is similar to the results for the Whittier section. Indeed many of these are large enough to be visible by eye in Fig. \ref{fig:map_socal}. There is some recognition of the strike direction parallel to the EFZ in the Coyote Mountain results, particularly for the 4 km scale. Therefore the faulting geometry appears to be more complex here and scale dependent.
        
        The final region of the Elsinore that we examine is the Yuha Desert. This area serves as the transition between the Elsinore and Laguna Salada systems and is underlain by the Paso Superior detachment fault \citep{fletcher_assembly_2014}. It was the site of extensive aftershock activity following the 2010 Mw 7.2 El Mayor-Cucapah earthquake, including the 2010 Mw 5.7 Ocotillo, California earthquake \citep{kroll_aftershocks_2013}. There also was a shallow Mw 6.5 slow slip event that occurred here as part of this sequence \citep{ross_aftershocks_2017}. The Yuha Desert area contains numerous fault traces orthogonal to the main trend of EFZ. Indeed our $K_{cyl}$ results corroborate this, with two modes with azimuthal separation of nearly 70 degrees. At the two largest scales, $t>100$ m and $r>1000$ m, the SE trending mode is stronger, whereas, for the smallest scale, the two modes are about equal in strength. There is no evidence for any significant deviation from vertical here, with the $r=2000$ m scale having a best estimate of $\delta=78^\circ$ NE.
        
        \section{Discussion and Conclusions}
        
        In this study we have outlined a new method for quantifying the average dip of fault zones using seismicity. Overall our results for southern California seismicity regions compare favorably with those of the SCEC CFM and other sources. While it is just one type of information, it is independent from that considered in the CFM. This study demonstrates the potential for using this method to augment existing CFM databases and ultimately improve upon the known geometrical properties of fault zones.
        
        Our primary findings for the major fault zones examined support the idea that the San Andreas and San Jacinto fault zones in southern California are dipping (at least in an average sense) toward the northeast. Most of the Elsinore fault zone is close to vertical, with the lone exception perhaps being the Whittier section at the northwest terminus of the fault zone near the LA Basin. Our findings suggest a progressive steepening of dip spatially, going from SAFZ in the northeast to EFZ in the southwest, which may provide clues as to the tectonic origins of this geometry. These conclusions are consistent with those of \citet{schulte-pelkum_tectonic_2020}.
        
        Our findings explicitly quantify anisotropy in seismicity at each length scale desired. There are hints of some changes with increasing length scale that may have broader implications about the tectonic history of the region. For example we found minor changes in dip with length scale that may suggest younger faults being formed in recent years may be inconsistent with the larger scale plate boundary founds surrounding them. Additional more detailed analysis is warranted for these cases to further substantiate these observations and possible implications.
        
        The method is not without limitations and these should be emphasized for further clarity on its usage. First, it should be understood that the cylindrical K function represents average properties over the window. Within the window, the properties may vary spatially, i.e. the seismicity may be viewed as an inhomogeneous point process. While the cylindrical K function is formulated under the assumption of stationarity, it can still provide useful information even if there are relatively mild deviations from this assumption. An important consequence of the lack of stationarity is that the results will depend on the spatial window chosen. They should be interpreted only for the specific region. This further implies that the results should not be extrapolated to regions outside of the spatial window. Another important limitation results from the "disc" geometry used to construct the cylindrical K function, which was chosen expressly with the purpose of detecting persistent planar features in seismicity. While not the focus of this study, other types of seismicity features, e.g. linear features, may not be detected with a disc geometry and would require alternatives.
        
        Location errors are the main source of measurement uncertainty in our calculations and their effects should be appropriately considered. The length scales of importance in our study are the values of $2r$, i.e. the diameter of the disc used in computing $K_{cyl}$. The values used are 1 km, 2 km, and 4 km. The seismicity catalog only included events with successful double-difference relocations and therefore the relative location error is the most important term to consider. For this catalog, 90\% of the events are estimated to have relative horizontal and vertical errors of 0.1 km, which is at least an order of magnitude smaller than the length scales considered. We therefore do not expect artifacts related to location errors.
        
		%% Will not be printed if anonymous option ON
		\section*{Acknowledgements}
	The author is grateful to the David and Lucile Packard Foundation for supporting this study through a Packard Fellowship.
		
		\section*{Data and code availability}
        
        The seismicity catalog used in this study is from \citet{hauksson_waveform_2012} and is publicly available from the Southern California Earthquake Data Center (https://scedc.caltech.edu) \citep{southern_california_seismic_network_southern_2013}. Maps were created with PyGMT \citep{pygmt_2023_7772533}.
		
		\section*{Competing interests}
		The authors have no competing interests.
		
		%% If the article is accepted, a separate bibfile must be uploaded along with the compiled manuscript, source file, and separate figure files.
		%% When available, DOI numbers must be provided for all references, including datasets and codes. 
            \bibliographystyle{unsrtnat}
		\bibliography{refs}
		
	\end{document}